\newcommand{\start}{\begin{document}}

\newcommand{\bye}{
\end{document}
\end}

\textwidth 6.5in 
	\oddsidemargin 0.0in
	\evensidemargin 0.0 in
	\textheight 8.5in
	\topmargin -.5in
	\topskip 0.0in

\newcommand{\nsfitp}[1]{
\begin{flushright}
NSF-ITP-{#1}
\end{flushright}}

\newcommand{\papertitle}[1]{
	\vskip 1in
	\begin{center}
	{\Large{\sc #1}}
        \end{center}
        \vskip .2in}

\newcommand{\by}[1]{
\begin{center}
{\large {#1}}\\
\vskip .1in}

\newcommand{\itp}{
{\footnotesize Institute for Theoretical Physics}\\
{\footnotesize University of California}\\
{\footnotesize Santa Barbara, CA 93106-4030 USA}\\
}

\newcommand{\itpline}{
{Institute for Theoretical Physics, University of California, 
Santa Barbara, CA 93106-4030 USA}}

\newcommand{\affiliation}[2]{
{$^#1$}{\footnotesize{#2}}\\}

\newcommand{\abs}{
\end{center}
\vskip .3in}

\newcommand{\bq}{\begin{equation}}
\newcommand{\eq}{\end{equation}}
\def\gele{{\mathop{\stackrel{>}{\scriptstyle <}}\nolimits}} 
\def\lege{{\mathop{\stackrel{<}{\scriptstyle >}}\nolimits}} 
\def\gtapprox{\mathrel{\mathpalette\overapprox>}} 
\def\ltapprox{\mathrel{\mathpalette\overapprox<}} 
\def\gtsim{\mathrel{\mathpalette\oversim>}} 
\def\ltsim{\mathrel{\mathpalette\oversim<}} 
\def\gtequal{\mathrel{\mathpalette\overequal>}} 
\def\ltequal{\mathrel{\mathpalette\overequal<}} 
\def\vev#1{{\left\langle{#1}\right\rangle}}
\def\pep#1{{\left\Vert{#1}\right\Vert}}
\def\pap#1{{\left({#1}\right)}}
\def\bab#1{{\left\lbrack{#1}\right\rbrack}}
\def\cac#1{{\left\lbrace{#1}\right\rbrace}}

\newcommand{\ann}{\sl{Ann.~Phys.\ }}
\newcommand{\cmp}{\sl{Commun.~Math.~Phys.\ }}
\newcommand{\np}{\sl{Nucl.~Phys.~A\ }}
\newcommand{\npa}{\sl{Nucl.~Phys.~A\ }}
\newcommand{\npb}{\sl{Nucl.~Phys.~B\ }}
\newcommand{\pla}{\sl{Phys.~Lett.~A\ }}
\newcommand{\plb}{\sl{Phys.~Lett.~B\ }}
\newcommand{\prp}{\sl{Phys.~Rep.\ }}
\newcommand{\pra}{\sl{Phys.~Rev.~A \ }}
\newcommand{\prb}{\sl{Phys.~Rev.~B \ }}
\newcommand{\prc}{\sl{Phys.~Rev.~C \ }}
\newcommand{\prd}{\sl{Phys.~Rev.~D \ }}
\newcommand{\pre}{\sl{Phys.~Rev.~E \ }}
\newcommand{\prl}{\sl{Phys.~Rev.~Lett. \ }}
\newcommand{\rmp}{\sl{Rev.~Mod.~Phys.\ }}

\newcommand{\grant}{This work was supported in part by the National 
Science Foundation under grant no.~PHY94-07194.\ }

\def\eg{{\it e.g.,\ }}
\def\ie{{\it i.e.,\ }}
\def\etal{{\it et al.}}
\def\etc{{\it etc.\ }}
\def\via{{\it via}}
\def\a{\alpha}
\def\be{{\beta}}
\def\ga{{\gamma}}
\def\de{{\delta}}
\def\eps{{\epsilon}}
\def\veps{{\varepsilon}}
\def\ze{{\zeta}}
\def\th{{\theta}}
\def\ka{{\kappa}}
\def\la{\lambda}
\def\si{{\sigma}}
\def\ups{{\upsilon}}
\def\om{{\omega}}
\def\G{{\Gamma}}                         
\def\D{{\Delta}}                         
\def\Th{{\Theta}}
\def\La{{\Lambda}}
\def\Si{{\Sigma}}
\def\Ups{{\Upsilon}}
\def\Om{{\Omega}}

\newcommand{\acknowledgement}{\section*{Acknowledgement}}



\def\endreferences{}
\def\refto#1{[#1]}
\def\refis#1{\parindent=0pt\hangindent=1cm\hangafter=1{[#1]\ }}

\def\[#1]{[\cite{[#1]}]}
\def\cite#1{[#1]}

\newcommand{\references}	    
  {\section*{References}	
   \frenchspacing \parindent=0pt \parskip=8pt plus 3pt 
	\everypar{\hangindent=\parindent}}


\catcode`@=11
\newcount\r@fcount \r@fcount=0
\newcommand{\refreset}{\global\r@fcount=0}
\newcount\r@fcurr
\immediate\newwrite\reffile
\newif\ifr@ffile\r@ffilefalse
\def\w@rnwrite#1{\ifr@ffile\immediate\write\reffile{#1}\fi\message{#1}}

\def\writer@f#1>>{}
\def\referencefile{
  \r@ffiletrue\immediate\openout\reffile=\jobname.ref%
  \def\writer@f##1>>{\ifr@ffile\immediate\write\reffile%
    {\noexpand\refis{##1} = \csname r@fnum##1\endcsname = %
     \expandafter\expandafter\expandafter\strip@t\expandafter%
     \meaning\csname r@ftext\csname r@fnum##1\endcsname\endcsname}\fi}%
  \def\strip@t##1>>{}}
\let\referencelist=\referencefile

\def\citeall#1{\xdef#1##1{#1{\noexpand\cite{##1}}}}
\def\cite#1{\each@rg\citer@nge{#1}} 

\def\each@rg#1#2{{\let\thecsname=#1\expandafter\first@rg#2,\end,}}
\def\first@rg#1,{\thecsname{#1}\apply@rg}	
\def\apply@rg#1,{\ifx\end#1\let\next=\relax
\else,\thecsname{#1}\let\next=\apply@rg\fi\next} 

\def\citer@nge#1{\citedor@nge#1-\end-}
\def\citer@ngeat#1\end-{#1}
\def\citedor@nge#1-#2-{\ifx\end#2\r@featspace#1 
  \else\citel@@p{#1}{#2}\citer@ngeat\fi}	
\def\citel@@p#1#2{\ifnum#1>#2{\errmessage{Reference
 range #1-#2\space is bad.}%
\errhelp{If you cite a series of references by the notation M-N, then M and
    N must be integers, and N must be greater than or equal to M.}}\else%
{\count0=#1\count1=#2\advance\count1 
by1\relax\expandafter\r@fcite\the\count0,%
  \loop\advance\count0 by1\relax
    \ifnum\count0<\count1,\expandafter\r@fcite\the\count0,%
  \repeat}\fi}

\def\r@featspace#1#2 {\r@fcite#1#2,}  
\def\r@fcite#1,{\ifuncit@d{#1}
    \newr@f{#1}%
    \expandafter\gdef\csname r@ftext\number\r@fcount\endcsname%
                     {\message{Reference #1 to be supplied.}%
                      \writer@f#1>>#1 to be supplied.\par}%
 \fi%
 \csname r@fnum#1\endcsname}
\def\ifuncit@d#1{\expandafter\ifx\csname r@fnum#1\endcsname\relax}%
\def\newr@f#1{\global\advance\r@fcount by1%
    \expandafter\xdef\csname r@fnum#1\endcsname{\number\r@fcount}}

\let\r@fis=\refis			
\long\def\refis#1#2#3\par{\ifuncit@d{#1}
   \newr@f{#1}%
   \w@rnwrite{Reference #1=\number\r@fcount\space is not cited up to 
now.}\fi%
  \expandafter\gdef\csname r@ftext\csname r@fnum#1\endcsname\endcsname%
  {\writer@f#1>>#2#3\par}}

\def\ignoreuncited{
   \def\refis##1##2##3\par{\ifuncit@d{##1}%
     \else\expandafter\gdef\csname r@ftext\csname 
    r@fnum##1\endcsname\endcsname%
     {\writer@f##1>>##2##3\par}\fi}}

\def\r@ferr{\endreferences\errmessage{I was expecting to see
 \noexpand\endreferences before now; I have inserted it here.}}
 \let\r@ferences=\references
 \def\references{\r@ferences\def\endmode{\r@ferr\par\endgroup}}

 \let\endr@ferences=\endreferences 

\def\endreferences
{\r@fcurr=0{\loop\ifnum\r@fcurr<\r@fcount\advance
        \r@fcurr by 1\relax\expandafter\r@fis\expandafter{\number\r@fcurr} %
        \csname r@ftext\number\r@fcurr\endcsname %
        \repeat}\gdef\r@ferr{}\global\r@fcount=0\endr@ferences}


\let\r@fend=\endpaper\gdef\endpaper{\ifr@ffile
\immediate\write16{Cross References written on 
[]\jobname.REF.}\fi\r@fend}

\catcode`@=12

\def\reftorange#1#2#3{$^{\cite{#1}-\setbox0=\hbox{\cite{#2}}\cite{#3}}$}

\citeall\refto		


\documentstyle[12pt]{article}
\baselineskip 36pt
\start
\papertitle{Law of addition in random matrix theory}
\by{A. Zee}
\itp
\abs
We discuss the problem of adding random matrices, which
enable us to
study Hamiltonians consisting of a deterministic term plus a
random term.
Using a diagrammatic approach and introducing the concept
of ``gluon
connectedness," we calculate the density of energy levels
for a wide class of
probability distributions governing the random term, thus
generalizing a
result obtained recently by Br\'ezin, Hikami, and Zee. The
method used
here may be applied to a broad class of problems involving
random
matrices.
\vskip .3in

\section{Introduction}

Some four decades ago, Wigner \refto{WIG} proposed studying the
distribution of energy levels of a random Hamiltonian given
by
\bq
H=\varphi
\label{eq:random}
\eq
where $\varphi$ is an $N$ by $N$ hermitean matrix taken
from the distribution
\bq
P(\varphi)={1 \over Z}{e^{{-N}
trV(\varphi)}}.
\label{eq:distribution}
\eq
with $Z$ fixed by
$\int d\varphi P(\varphi)=1$.
This problem has been studied intensively by Dyson, Mehta,
and others
over the
years \refto{POR,MEH,JER}.
Two years ago, Br\'ezin and Zee discovered that,
remarkably, while the
density of eigenvalues depends \refto{BIPZ} on $V$,
the correlation between the density of eigenvalues,
when suitably scaled, is
independent \refto{BZ1} of $V$. This universality was also
obtained
earlier and independently by Ambj\o rn and collaborators
\refto{amb}. Since
then, it  has been clarified and
extended by other authors \refto{bee, eyn, for}, studied
numerically \refto{koba}, and furthermore, shown to hold
even when the
distribution (\ref{eq:distribution}) is generalized to a
much broader class of distributions \refto{BZ2}. We expect
that the
discussion to be given below will hold also for this broader
class of
distributions, but for the sake of simplicity we will not work
this
through here.

In recent work \refto{BZ3},  Br\'ezin and Zee have
generalized
this Wigner problem to the case of a Hamiltonian given by
the sum of a deterministic term and a random term
\bq
H= H_0 + \varphi
\label{eq:det}
\eq
Here $H_0$ is a diagonal matrix
with diagonal
elements $\epsilon_i$, $i=1,2,...N$, and $\varphi$ a random
matrix taken from the ensemble (\ref{eq:distribution}). For
the
Gaussian case, namely with
$V(\varphi)={1\over2}\varphi^2$, Pastur \refto{PAS} has
long ago determined the density of eigenvalues. The work
described in
\refto{BZ3} went beyond Pastur's work in that the
correlation function between the density of eigenvalues in
the Gaussian
case was also determined. More recently, in a work with
Br\'ezin and
Hikami \refto{BHZ}, we managed to determine the density of
eigenvalues for $V(\varphi)={1\over2}\varphi^2 + g
\varphi^4$ to all orders in $g$. The
correlation function was also computed, but only to first
order in $g$.

This problem of ``determinism plus chance" may be
regarded
as a generic problem in physics, and as such represents a
significant generalization of Wigner's problem. For
example, consider an electron moving in a magnetic field
and scattering off
impurities. We note that these ``deterministic plus random"
problems
are considerably more difficult than the purely random
problems defined in (\ref{eq:random}) and
(\ref{eq:distribution}). A standard
approach to
solving the purely random problem involves diagonalizing
the random matrix $\varphi$ and then use orthogonal
polynomials to disentangle the resulting expression. Clearly,
in (\ref{eq:det}) we cannot diagonalize $\varphi$ without
un-diagonalizing $H_0$ and thus the orthogonal polynomial
approach
fails.

In this paper, we point out that the problem given in
(\ref{eq:det}) is
a special case of a broader class of problems involving the
addition of random matrices. The deterministic Hamiltonian
$H_0$ may in turn be replaced by a random Hamiltonian.
Indeed, a deterministic matrix is but a special case of a
random matrix. We will extend the work of Br\'ezin, Hikami,
and
Zee \refto{BHZ} and determine the density of eigenvalues
for the Hamiltonian given in (\ref{eq:det}) for an arbitrary
$V$.

Our work is inspired by recent advances in the
mathematical literature involving the theory of
non-commutative probability and operator algebra
\refto{voi,hag,waterloo}. A number of physicists have
already
brought these advances to the attention of the physics
community \refto{douglas, gross, li}. While our work is thus
inspired, we will not be using the mathematical approach
given in \refto{voi,hag,waterloo}, but instead will be based
on the diagrammatic approach developed in \refto{BZ3} and
subsequent work \refto{lattice, BHZ}.

\section{Adding Random Matrices}

Consider a
Hamiltonian given by
\bq
H=\varphi_1 +\varphi_2
\label{eq:ham}
\eq
with the matrices $\varphi_{1,2}$ taken from the
probability distribution
\bq
P(\varphi_1, \varphi_2)={1 \over Z}{e^{{-N}
tr [V_1(\varphi_1) + V_2(\varphi_2)]}} \equiv
P_1(\varphi_1)P_2(\varphi_2).
\label{eq:distributionproduct}
\eq
Notice that the probability distribution factorizes. (This is
known as ``free" in the mathematical literature.) The
problem defined in
(\ref{eq:det}) represents a special case. Previously, with
D'Anna and with
Br\'ezin we have studied the problem given in
(\ref{eq:ham}) but for the
more difficult case \refto{bz4, danna} in which
$P(\varphi_1, \varphi_2)$ contains terms linking
$\varphi_1$ and $\varphi_2$. Indeed, a detailed
determination of the correlation function over all ``distance
scales" is a non-trivial problem even for Gaussian
distributions \refto{danna}. The discussion in this paper
goes
through
precisely because $\varphi_1$ and $\varphi_2$ do not
couple to each other
in $P(\varphi_1, \varphi_2)$.

Let us now mention a few necessary definitions.
Define the Green's function
\bq
G(z)\equiv \left\langle{{1\over N}tr{1\over z-
H}}\right\rangle = \int \int
d\varphi_1 d\varphi_1P(\varphi_1, \varphi_2) {{1\over
N}tr{1\over
z-(\varphi_1 + \varphi_2)}}
\label{eq:1.4}
\eq
The density of eigenvalues is then given by
$\rho(\mu)=\left\langle{{1\over N}tr\delta(\mu-
H)}\right\rangle=-
{1\over\pi} {\rm{Im}}
G(\mu+i\epsilon).$
In this paper we focus on the density of eigenvalues,
leaving the correlation for a future work. Note that the
factors of $N$ are chosen in our definitions such that the
interval over which $\rho(\mu)$ is non-zero is
finite (\ie of order
$N^0$) in the large $N$ limit.

We may regard the distribution
(\ref{eq:distributionproduct}) as
defining a $(0+0)$-dimensional field theory. The Feynman
diagram
expansion is then simply obtained by expanding $G(z)$ in
inverse powers of $z$ and doing the integrals in
(\ref{eq:1.4}). As
explained in \refto{BZ3}, it is useful
to borrow the terminology of large $N$
quantum chromodynamics \refto{thoo}  from the
particle physics literature,  and speak of quark and gluon
lines. See figure (1) for a graphical representation. (It is of
course not
necessary to use this language, and readers not familiar
with this language
can simply think of the diagrams as representing the
different terms one
encounters in doing the integral in (\ref{eq:1.4})). The
quark propagator
simply
comes from the explicit factor of $z$ in (\ref{eq:1.4}) and is
represented
by a single line and given by $1\over z$. The quadratic
terms in
$V_1(\varphi_1)$ and $V_2(\varphi_2)$ determine the
gluon propagator,
represented by double lines.
Here, in a minor departure from large $N$
quantum chromodynamics we
have two types of gluons, corresponding to $\varphi_1$ and
$\varphi_2$. The
gluon propagators are proportional to
\bq
\left\langle{\varphi^i_{\alpha j}\varphi^k_{\beta
l}}\right\rangle
 \propto \delta_{\alpha \beta}\delta^i_l\delta^k_j{1\over
N}.
\label{eq:2.3}
\eq
The non-Gaussian terms in $V_1(\varphi_1)$ and
$V_2(\varphi_2)$
describe
the interaction between the gluons.

The reason that we can solve this problem is because, while
the two types
of gluons
have arbitrarily complicated interactions among
themselves, they do not interact with each other. Note that
while the gluons both interact with the quark, our
problem is such that we do not have to include quark loops
and thus the quark does not induce interaction between the
two gluons. This is clear from the definition of our problem.
Another way of saying this is to note that the Green's
function
may be represented, by using the replica trick, as
\bq
G(z) = lim_{n \rightarrow 0} \int D\psi^{\dagger} D\psi
D\varphi
P(\varphi)\psi_1^{\dagger} \psi_1e^{-\sum_{\alpha=1}^n
\psi_{\alpha}^{\dagger} (z-\varphi)\psi_{\alpha}}
\label{eq:field}
\eq
Note that in this language the $\psi$'s represent the quark
fields and
$\varphi$ the gluon fields. The interaction between gluon
and quarks are
given by $ \psi_{\alpha}^{\dagger}\varphi\psi_{\alpha}$.
(Color indices
are
suppressed here.) The interaction of the gluons with each
other is
determined by $P(\varphi)$.
Since internal quark loops are proportional to the number
of replicas $n$, they vanish in the $n\rightarrow 0$ limit.

Let us then calculate the Green's function, which as usual
can be written
as (see figure 2)
\bq
G(z)={1\over z-\Sigma(z)}
\label{eq:ipi}
\eq
in terms of the one-particle  irreducible self energy
$\Sigma^i_j(z)= \delta^i_j \Sigma(z)$.  The self-energy is
then
determined by the set of diagrams in figure (3) with the
corresponding
equation
\begin{eqnarray}
\Si(z)
&=& <{1\over N} tr \varphi_1>_{gc}+<{1\over N} tr
\varphi_1^2>_{gc} G(z) + <{1\over N} tr
\varphi_1^3>_{gc}G(z)^2 +
.... + (1 \leftrightarrow 2) \nonumber\\
&=& \sum_{k=1}^{\infty} <{1\over N}
tr\varphi_1^k>_{gc}G(z)^{k-
1}  + (1 \leftrightarrow 2) \nonumber\\
&=& {1\over G} \left[ <{1\over N} tr {1\over 1-\varphi_1
G}>_{gc}
-1\right] + (1 \leftrightarrow 2) \nonumber\\
&=& {1\over G} \left[ {1\over G} G_{gc1}({1\over G})-
1\right] +
(1 \leftrightarrow 2). \nonumber\\
\label{eq:cent}
\end{eqnarray}
In order to write this equation, we have to invoke the
factorization of
$P(\varphi_1, \varphi_2)$, which tells us that the two kinds
of gluons do
not interact, and the large $N$ limit, which tells us that the
two kinds of
gluon lines cannot cross.

We are led to
introduce in (\ref{eq:cent}) the notion of ``gluon
connectedness,"
denoted by ``gc"
henceforth. The necessity for this notion is illustrated by
the shaded
blob describing the interaction of the gluons in figure (3d):
it should not
include the diagram shown in figure (4): this class of
diagrams
is already included in figure (3b). In other words, a gluon
connected blob
with $k$ external gluon lines is such that it cannot be
separated into two
blobs, with  $k_1$ gluon lines and $k_2$ gluon lines
respectively, (with
$k_1+k_2=k$ of course). In the last line we have defined the
``gluon
connected Green's function"
\bq
G_{gc1}(z)=<{1\over N} tr {1\over {z-\varphi_{1}}}>_{gc}
\label{eq:ggc}
\eq
and similarly
$G_{gc2}(z)$. The operations implied in (\ref{eq:ggc}) are
clearly
allowed since $< {1\over N} tr \cdot >_{gc}$ is a linear
operation. Note also that we have not assumed that
$V_\alpha$ is an even
function of its argument. In particular, we include a possible
tadpole term indicated by $ <{1\over N} tr
\varphi_{\alpha}>_{gc}$ in (\ref{eq:cent}).

We should emphasize that the shaded blobs include
interactions between
gluons to all orders. It is very complicated, if not hopeless, to
calculate
these blobs in terms of $V_1$ and $V_2$, but fortunately, as
we will show
below, we do not have to calculate them explicitly.
In our previous papers,
we regarded the cubic, quartic, and so on, terms in
$V_\alpha$ as  interactions and proceeded to calculate the
Green's
function and correlation function in terms of
the various coupling constants. We follow a different
strategy here, and try to express the Green's function
$G(z)$ directly in terms of $G_1(z)$ and $G_2(z)$ where
\bq
G_{\alpha}(z) \equiv <{1\over N} tr {1\over  z -
\varphi_{\alpha} }>
\label{eq:alpha}
\eq
for $\alpha =1,2$ are the Green's functions for two
separately and purely random problems. (The average in
(\ref{eq:alpha})
is performed with the distribution
$P_\alpha(\varphi_\alpha)={1\over
Z_\alpha}e^{-NtrV_\alpha(\varphi_\alpha)}$ of course.) In
this way, we
attempt to bypass having to deal with $V_1$ and $V_2$
altogether.

To see how to do this, let us go back to the simpler problem
defined by
(\ref{eq:random}) and (\ref{eq:distribution}).
Following the same diagrammatic analysis leading to
(\ref{eq:cent}) we find
that the
Green's function $G(z)$ and self energy $\Si(z)$ for this
simpler problem
are related by
\bq
\Si(z) = {1\over G} \left[ {1\over G} G_{gc}({1\over G})-
1\right]
\label{eq:cent1}
\eq
with, evidently,
\bq
G_{gc}(z) \equiv <{1\over N} tr {1\over {z-\varphi}}>_{gc}
\label{eq:def}
\eq
Combining (\ref{eq:def}) and (\ref{eq:ipi}) we find
\bq
{1\over G^2} G_{gc}({1\over G}) = z
\label{eq:zip}
\eq
For the sake of convenience, we may, with due respect to
Green, somewhat
fancifully define a
``Blue's function" by
\bq
B(z) \equiv {1 \over z^2} G_{gc}({1 \over z})
\label{eq:blue}
\eq
Thus, we learn from (\ref{eq:zip}) that the Blue's function is
the functional
inverse of the
Green's function
\bq
B(G(z))=z
\label{eq:inverse}
\eq
From the normalization of the probability distribution
$P(\varphi)$ we
obtain trivially the ``sum rule" $G(z)\rightarrow{1\over z}$
as $z\rightarrow\infty$, thus
implying that the Blue's function $B(z)\rightarrow{1\over
z}$ as $z\rightarrow0$.

Let us now go back to the more involved problem
defined by (\ref{eq:ham}). First, we define for $\alpha=1,2$
two Blue's
functions $B_{\alpha}$ as the
functional inverse
of $G_{\alpha}$ respectively. We now see that (\ref{eq:cent}),
when combined
with
(\ref{eq:ipi}), says simply that
\bq
z+{1\over G}= B_1(G) + B_2(G)
\label{eq:add}
\eq
Thus, the law of addition for the Blue's function is given by
\bq
B_{1+2}(z)=B_1(z)+B_2(z)-{1\over z}
\label{eq:addlaw}
\eq
This equation tells us how to obtain the Blue's function
associated with
$\varphi_1 + \varphi_2$
from the Blue's functions associated with
$\varphi_1$ and
$\varphi_2$.

The procedure for determining the Green's function and
hence the density of
eigenvalues of the problem defined by (\ref{eq:random})
and
(\ref{eq:distribution}) is then
as follows: given the Green's functions $G_1$ and $G_2$,
determine the
corresponding Blue's functions $B_1$ and $B_2$ by
functionally inverting
$G_1$ and $G_2$ respectively,  calculate
$B_{1+2}$
according to (\ref{eq:addlaw}), then determine the
functional
inverse of  $B_{1+2}$
to find the desired Green's function $G(z)$.

Let us remark briefly on the connection to the mathematical
literature.
Voiculescu \refto{voi} has introduced the ``$R$-transform."
It
turns out that
the $R$ function discussed by mathematicians is simply
related to $B$ by
$B(z)={1\over z} + R(z)$. In fact, we see that the
Dyson-Schwinger equation
(\ref{eq:ipi}) when combined with (\ref{eq:inverse}) gives
simple
$B(G(z))={1\over G(z)} +
\Si(z)$. Thus, the $R$ function of the mathematicians is
nothing but the
self-energy $\Si$ of the physicists expressed in terms of
different
arguments:
\bq
R(G(z))=\Si(z)
\label{eq:math=phys}
\eq

\section{Addition Algorithm at Work}

Let us now proceed by building up from a few simple
examples. In the most trivial case, $\varphi$ is
not random at all, but fixed to be a constant $c$ times the
unit matrix. Then from the Green's function $G(z)= {1\over
z-c}$ we find the Blue's function $B(z)=c+{1\over z}$. For a
slightly less trivial example, let $\varphi$ be a diagonal
matrix with matrix elements given by $\epsilon_i$ with $i=1,
..., N$. The Green's function is given by
\bq
G(z)={1\over N}\sum_i{1\over z-\epsilon_i}
\label{eq:Gfordet}
\eq
Then the corresponding Blue's function is determined
by
\bq
{1\over N}\sum_i{1\over B(z)-\epsilon_i}=z
\label{eq:bluedet}
\eq
Next, let $P(\varphi)$ be  Gaussian (that is, $V(\varphi)=tr
{1\over2}\varphi^2$). Then as is well known (see for
example
\refto{BZ3}), the
Green's function is determined by
\bq
z=G(z)+{1\over G(z)}
\label{eq:gauss}
\eq
In other words,
\bq
G(z)={1\over 2}(z-{\sqrt {z^2-4}})
\label{eq:formforG}
\eq
Substituting $z\rightarrow B$ into (\ref{eq:gauss}), we
obtain immediately
that \refto{addgauss}
\bq
B(z)=z+{1\over z}
\label{eq:gaussb}
\eq
Thus, in this simple case, the Blue's function $B$, which of
course
contains the same information as the Green's function $G$,
actually has a
simpler form than $G$.

Now we are ready to do our first non-trivial problem.
Consider the problem defined in (\ref{eq:det}). Since we
know from
(\ref{eq:bluedet}) and (\ref{eq:gaussb}) the Blue's functions
corresponding
to the two terms in the Hamiltonian, we learn immediately
from (\ref{eq:addlaw}) the Blue's function for $H$:
\bq
B_{1+2}(z)=B_1(z)+z+{1\over z}-{1\over z}
=B_1(z) + z
\label{eq:fullblue}
\eq
with $B_1$ determined by (\ref{eq:bluedet}) with the
substitution
$B \rightarrow B_1$.
The desired Green's function $G(z)$ is now determined by
solving for the functional inverse of the function $B_{1+2}$,
that is, by the equation
\bq
B_{1+2}(G)=z
\label{eq:iii}
\eq
or equivalently, upon using (\ref{eq:fullblue}),
\bq
B_1(G)=z - G
\label{eq:iv}
\eq

Let us now evaluate the two sides of this equation with the
function $G_1(\cdot)$.
Since $G_1(B_1(G(z)))=G(z)$ we obtain
immediately
\bq
G(z)=G_1(z-G(z))
\label{eq:past}
\eq
Noting that $G_1$ is given by (\ref{eq:Gfordet}) with the
substitution $G
\rightarrow G_1$, we have
\bq
G(z)={1\over N}\sum_i{1\over z-\epsilon_i-G(z)}
\label{eq:Gfordddd}
\eq
precisely the classic result of Pastur which was obtained
diagrammatically in \refto{BZ3}.

After these simple exercises, we can now immediately go on
and solve the general version of the problem defined in
(\ref{eq:det}): find the density of energy levels of a
Hamiltonian
given by $H=H_0 + \varphi$ with $\varphi$ drawn from the
general distribution (\ref{eq:distribution}). With a slight
shift in
notation, let us call the Blue's function associated with $H_0$
and with $\varphi$ respectively $B_0$ and $B_2$. Then the
Blue's function associated with $H$ is given by $B(z)=B_0(z)
+B_2(z) - {1\over z}$. Substituting in this equation $z
\rightarrow G(z)$
(where $G(z)$ is the unknown Green's function associated
with $H$), we find immediately that
\bq
B_0(G)=z+{1\over G}-B_2(G)
\label{eq:daf}
\eq
Anticipating the next step, we define
\bq
\Si(z)=B_2(G(z))-{1\over G(z)}
\label{eq:adef}
\eq
With this definition, we write (\ref{eq:daf}) as
\bq
B_0(G)=z-\Si(z)
\label{eq:daf2}
\eq
Let us now evaluate both sides of (\ref{eq:daf}) with the
Green's
function $G_0(\cdot)$ associated with $H_0$. We find
instantly
that
\bq
G_0\left( B_0(G(z))\right) = G(z)=G_0\left( z-\Si(z)\right)
= {1\over N} \sum_i {1\over z-\epsilon_i-\Si(z)}
\label{eq:one}
\eq
Let us repeat this trick: rewrite (\ref{eq:adef}) as
$B_2(G(z))=\Si(z)+{1\over G(z)}$ and evaluate both sides
with the function $G_2(\cdot)$. We obtain
\bq
G(z)=G_2(\Si(z) + {1\over G(z)}).
\label{eq:other}
\eq

These two equations, (\ref{eq:one}) and (\ref{eq:other}),
allow us to
determine the two unknown functions $G(z)$ and $\Si(z)$,
provided we know the Green's function $G_2(z)$. But what
is the Green's function $G_2(z)$? It is just the Green's
function associated with the random matrix $\varphi$
drawn from the general distribution (\ref{eq:distribution}).
But this was
obtained by Br\'ezin et al \refto{BIPZ} almost
twenty years ago. These authors told us that (for $V(z)$ an
even polynomial for the sake of notational simplicity)
\bq
G_2(z)={1\over2}[V'(z)-P(z){\sqrt {z^2 - a^2}}]
\label{eq:bipz}
\eq
Here $V'(z)\equiv {dV \over dz}$, $P(z)$ is a polynomial, and
$a$ determines the endpoints of the spectrum of
eigenvalues. The quantities $P(z)$ and $a$ are determined
\refto{foot}
by the ``sum rule" that $G_2 \rightarrow {1\over z}$ as $z
\rightarrow \infty$.

In summary, and repeating various equations for clarity,
we have obtained
the following result. For a Hamiltonian of the form $H=H_0 +
\varphi$ with
the random matrix $\varphi$ drawn from an arbitrary
distribution defined by
$V(\varphi)$ (taken to be even for simplicity), we can
determine the Green's function $G(z)$ and hence the
density of
eigenvalues by solving simultaneously the two equations
\bq
G(z)= {1\over N} \sum_i {1\over z-\epsilon_i-\Si(z)}
\label{eq:one'}
\eq
and
\bq
G(z)=G_2(\Si(z) + {1\over G(z)}).
\label{eq:other'}
\eq
where the function $G_2$ is given by (\ref{eq:bipz}). In
general, for an
arbitrary set of $\epsilon_i$'s and a non-Gaussian $V$, these
equations can
only be solved numerically.

It is clearly of some notational benefit to give the
combination appearing
in (\ref{eq:other'}) a name: $\si(z) \equiv \Si(z) + {1\over
G(z)}$. We can
then
simplify
(\ref{eq:other'}) slightly to \refto{passing}
\bq
P^2(\si)(\si^2 - a^2)=(V'(\si)-2G)^2
\label{eq:sigma}
\eq
Thus, we can use (\ref{eq:sigma}) to determine $\si$, and
hence $\Si$, in
terms
of $G$. Plugging this into (\ref{eq:one'}) then gives us an
equation for $G$.

As mentioned earlier, Br\'ezin, Hikami,
and Zee \refto{BHZ} recently used the equation of motion
method and a
detailed
diagrammatic analysis to determine the Green's function
for the
problem in
(\ref{eq:det}) with the distribution defined by
$V(\varphi)={1\over
2}\varphi^2 + g \varphi^4$. It is straightforward, although
slightly
tedious, to verify that for this simple
case, (\ref{eq:sigma}) reduces to equation (4.15) in
\refto{BHZ}. The analysis
given here is considerably simpler.

\section{Energy Bands}

The function $B$ contains the same amount of
information as $G$, since one function is the inverse of the
other.
As an example, consider the problem of determining the
endpoints of the energy spectrum. Near the endpoint, call it
$a$,
the density of states $\rho(\mu)$, that is, the imaginary
part of
$G(z)$, vanishes generically like $(a-\mu)^{1\over 2}$. Thus,
the
endpoint is determined by the equation
\bq
{dG\over dz} |_{z=a} = \infty
\label{eq:dog}
\eq
Now consider the defining equation for $B$: namely
$G(B(w))=w$.
Differentiating, we
obtain
\bq
{dG\over dB} {dB \over dw} =1
\label{eq:cat}
\eq
Thus, we obtain an alternative equation for determining the
endpoints $a$ of the energy spectrum of a random
Hamiltonian:
solve the equation
\bq
{dB \over dw} |_{B=a} = 0
\label{eq:mouse}
\eq
Let us now apply these simple considerations to determine
the endpoints of the spectrum of a Hamiltonian of the form
$H=H_0 + \varphi$.

As remarked above, with an arbitrary $H_0$ (that is, an
arbitrary set of $\epsilon_i$'s) and a general $V$,
we can hardly expect to solve
(\ref{eq:one'}) and (\ref{eq:other'}) analytically. Let us
retreat to the
case in which
$\varphi$ is
taken from a Gaussian distribution, so that $\Sigma(z)$ in
(\ref{eq:one'})
can be replaced by $G(z)$. (In other words, the solution of
(\ref{eq:other'}) is simply $\Si(z)=G(z)$.) Thus, we have to solve \bq
G(z)={1\over N}\sum_i^N {1\over z-\epsilon_i-G(z)}
\label{eq:eqforg}
\eq
We will be interested in the case in which the Hamiltonian
$H_0$
exhibits degeneracy. Let $K$ be the number of distinct
$\epsilon_i$'s (with $K \leq N$ of course.)
To determine $G$ we have to solve a polynomial equation of
degree $K+1$.

As explained above, in general, given an equation
determining $G(z)$, we simply
substitute $z \rightarrow B(w)$ and $G(z) \rightarrow w$
into that
equation to obtain the equation for determining $B(w)$.
Thus,
in the
present example, the function $B$ is determined by
\bq
w={1\over N}\sum_i^N {1\over B(w) -\epsilon_i-w}
\label{eq:bach}
\eq
We note that this is a polynomial equation of degree $K$,
that is,
of degree one lower than the equation for determining $G$.
In particular, for the case of $K=2$, while we have to solve a
cubic
equation to determine $G$, we only have to solve a
quadratic
equation to determine $B$!

We will now exploit this fact and study the
Hamiltonian  $H = H_0
+
\varphi$ for $K=2$, in other words, with no loss of
generality we take the
symmetric
case with
the deterministic piece
\bq
H_0 = \left( \begin{array}{cc}
\epsilon & 0 \\
0 & -\epsilon \\
\end{array}
\right)
\label{deter}
\eq
and take $\varphi$ to be a Gaussian random perturbation.
We can
imagine
some possible physical applications of this Hamiltonian. For
instance, consider electron scattering on impurities in a spin
dependent quantum Hall fluids. In the absence of impurity
scattering
we
have a spin up Landau level separated in energy
from the
spin down Landau level, with a Zeeman splitting of
$2\epsilon$.
Impurity scattering is represented schematically by
$\varphi$. A
model not precisely of this type, but in the same spirit, was
proposed by Hikami, Shirai, and Wegner [8] and has been
studied by a number of authors [9,10].  The reader can
no doubt concoct
other
possible situations represented, at least schematically, by a
Hamiltonian
of
the type considered here.

For $\epsilon=0$, the density of states of $H$ is given by the
familiar semi-circle law  $\rho(\mu)= {1\over
2\pi}\sqrt{4-
\mu^2}$ which we obtain easily from (3). Note that the half
width of the
spectrum is equal to $2$. In the
other limit,
$\epsilon >> 2$, the density of
states
decompose into two pieces. At some critical value
$\epsilon_c$ the two disjoint pieces in the density of states
touch
and merge into one piece. An interesting question is
whether
$\epsilon_c$ is larger or smaller than 2. A simple physical
argument based on level repulsion would suggest that
$\epsilon_c < 2$.

The equation (\ref{eq:eqforg}) for $G$ becomes
\bq
2G={1\over z-\epsilon-G} + {1\over z+\epsilon-G}
\label{eq:2G}
\eq
While an explicit solution of this cubic equation may be
written down, it
is rather
unwieldy. In contrast, as just explained,
$B$ satisfies the quadratic equation
\bq
z B(z)^2 - (2 z^2+1)B(z) + z^3 + (1-\epsilon^2) z =0
\label{eq:zb2}
\eq
whose solution can of course be immediately written down.

We would now like to determine the endpoints of the
spectrum of $H$ using
the algorithm we developed above. We employ the following
simple
trick. Define the operator ${d \over dz} |
$ with the vertical bar indicating that the derivative should
not act
on $B$. This operator is clearly useful  in light of
(\ref{eq:mouse}).
We now act on  (\ref{eq:zb2}) with the operator $(z {d \over
dz} |
-1)$. The  net effect of this
operator is to simply replace $z^n$ in (\ref{eq:zb2}) by $(n-
1)z^n$. We
thus
obtain a linear equation for $B$:
\bq
(2 z^2 -1)B - 2 z^3 =0
\label{eq:2z2}
\eq
We solve this equation and (\ref{eq:zb2}) (or somewhat
more simply, the
equation obtained by acting with
${d
\over dz} |
$ on (\ref{eq:zb2})) simultaneously for $z$ and $B$. The
values of
$B$
thus obtained are in fact the values of $a$ that we are
trying to
determine.

Proceeding in this way, we find easily that the four values of
$a$ are
given by
\bq
a_+ = {1\over 2{\sqrt 2} \epsilon} { (4 \epsilon^2 - 1 + {\sqrt
{8
\epsilon^2 +1}})^{3\over 2} \over {\sqrt {8 \epsilon^2 +1}} -
1}
\label{eq:a+}
\eq
\bq
a_- = {1\over 2{\sqrt 2} \epsilon} { (4 \epsilon^2 - 1 - {\sqrt{
8
\epsilon^2 +1}})^{3\over 2} \over {\sqrt {8 \epsilon^2 +1}} +
1}
\label{eq:a-}
\eq
and $-a_+$ and $-a_-$.

As a first check, we take the limit $\epsilon \rightarrow 0$.
We find
that as expected, two of the $a$'s, namely $\pm a_-$ become
complex and hence unphysical. Indeed, $a_+ \rightarrow 2$,
and we recover Wigner's classic result for a Gaussian
random Hamiltonian.

In the opposite limit with $\epsilon$ large, we find
\bq
a_+ = \epsilon (1+{{\sqrt 2}\over \epsilon} + ...)
\label{eq:yellow}
\eq
and
\bq
a_- = \epsilon (1 - {1\over {\sqrt 2}\epsilon} + ...)
\label{eq:orange}
\eq
Thus, in the density of states the width of each of the two
disjoint pieces, or bands, is equal to ${\sqrt 2}+{1\over
{\sqrt 2}} =
{3\over
{\sqrt 2}}$. Thus, even when the two bands are far apart,
the width
is reduced compared to the Gaussian bandwidth by
\bq
{{\rm bandwidth}\over{{\rm Gaussian}\,{\rm bandwidth}}}
= {3\over4{\sqrt 2}}
\label{eq:brown}
\eq

Finally, the phase transition when the two disjoint bands
merge
into one occurs when $a_-$ vanishes and becomes complex.
This
occurs at
\bq
\epsilon_c =1
\label{eq:maroon}
\eq
As expected, $\epsilon_c$ is less than the
unperturbed Gaussian half-width of 2. Interestingly, level
repulsion
reduces the naive expectation by precisely a factor of two.

Note that for the purpose of this discussion we never had to
solve
for $B$ explicitly, but the solution of (\ref{eq:zb2}) is of course easy
enough to
write
down
for the
records:
\bq
B(z)={2 z^2 + 1+ {\sqrt{1+ 4 \epsilon^2 z^2}}\over 2z}
\label{eq:pink}
\eq
The function $B(z)$ has two branch cuts starting at
$z_c=\pm
{i\over 2\epsilon}$. We check easily that the addition law
in (\ref{eq:addlaw}), giving $B(z)=B_0(z)+z$, with $B_0$ the
$B$ function
associated with
$H_0$ of course,
produces the same result.

Now that we have discussed a specific example, let us now go
back to the
general case and write (36) as
\bq
z=\int d\mu {\sigma(\mu) \over
 B-\mu-z}
\label{eq:disp}
\eq
where $\sigma(\mu)={1\over N} \sum_i \delta(\mu-
\epsilon_i)$ is the density of states of the deterministic
Hamiltonian $H_0$. Again, with no loss of generality, we
have set $m^2 =1$ in the Gaussian distribution governing
$\varphi$.
The endpoint $a$ of the spectrum of $H$ is given by the
value of
$B$ that simultaneously solves (\ref{eq:disp}) and the
equation
\bq
1 = \int d\mu {\sigma(\mu) \over
 (B-\mu-z)^2
}
\label{eq:unit}
\eq
Note that as explained before (\ref{eq:unit}) is obtained by
acting with
the operator ${d\over dz} |$ on (\ref{eq:disp}).

Given any $\sigma(\mu)$ we can thus determine in
principle the endpoints of
the energy spectrum of $H$.
We can for example consider  the limit of large
randomess,
that is, when the spectrum of $H_0$ is small compared to
the
scale of the randomness. Expanding the denominators in
(\ref{eq:disp})
and (\ref{eq:unit}) and
defining $<\epsilon^{2k}>\equiv\int d\mu \sigma(\mu)
\mu^{2k}$ (assuming for
simplicity that $\sigma(\mu)$ is even) we obtain
\bq
z={1\over (B-z)} + {<\epsilon^{2}>\over (B-z)^3}
+{<\epsilon^{4}>\over
(B-z)^5}+......
\label{eq:purple}
\eq
and
\bq
1= {1\over (B-z)}+{3<\epsilon^{2}>\over (B-
z)^4}+{5<\epsilon^{4}>\over
(B-z)^6}+......
\label{eq:silver}
\eq
In the limit $H_0 \rightarrow 0$ or large randomness, we
have $z={1\over
B-z}=\pm 1$ (and thus the endpoints $a=\pm 2$, as
expected.) It is easy to
solve these equations to any desired power. We obtain to
$O(H_0^4)$ that the width of the spectrum is given by
\bq
a=2+<\epsilon^2>-{9\over 4}<\epsilon^2>^2+<\epsilon^4>+....
\label{eq:gold}
\eq
A simple check shows that indeed, if we expand $a_+$ in
(\ref{eq:a+})
we find
\bq
a_+= 2+\epsilon^2-{5\over 4}\epsilon^4 + .....
\label{eq:white}
\eq
This is consistent with (\ref{eq:gold}) since in this simple
model
$<\epsilon^2>^2=<\epsilon^4>=\epsilon^4$.

\section{Analytic Structure}

For the problem studied here $B$ is simpler than $G$, but in
other
problems $B$ is unfortunately more complicated. For
instance, for a much studied class of probability
distribution, the so-called trace class,  defined by
(\ref{eq:distribution}), and with $V$ taken to be an even
polynomial for
simplicity,
$G(z)$ is given by \refto{BIPZ}, as mentioned earlier,
\bq
G(z)= {1\over 2}\left[ V'(z)- P(z)\sqrt {z^2 - a^2}\right]
\label{eq:chablis}
\eq
where the polynomial $P$ and the endpoint $a$ are
determined
completely by the condition $G(z) \rightarrow 1/z$ as $z
\rightarrow \infty$. Applying our rule of substituting $G \rightarrow w$
and $z \rightarrow B$, we obtain the equation
\bq
2w=V'(B)-P(B)\sqrt {B^2 - a^2}
\label{eq:cabernet}
\eq
which determines $B$.

For $V$ of degree $2s$ it is easy to see that $B$ satisfies a
polynomial
equation of degree $2s-1$. Indeed, moving $V'(B)$ in
(\ref{eq:cabernet}) to
the left hand side and squaring, we can rewrite
(\ref{eq:cabernet}) as
\bq
wV'(B)-w^2 - Q(B)=0
\label{eq:merlot}
\eq
where
$ 4Q(B) \equiv V'(B)^2 - P(B)^2 (B^2-a^2)$ is a polynomial
of degree $2s-1$
in $B$. (This is most easily seen by noting that $G(B)
\rightarrow {1/
B}$ for large $B$ by definition.)

We would like to study the analytic structure of  $B$ but we
have not been
able to make general and complete statements. One limited
statement applies
to
the trace class just discussed. It is easy to show then that
the branch cuts of $B(w)$ is of the square root type. Let
$B(w)$ has a cut starting at $w*$ and write $B*=B(w*)$. At
the tip of
the cut,
\bq
{dB(w)\over dw}|_{w*}=\infty
\label{eq:chardonnay}
\eq
or
\bq
{dG(B)\over dB}|_{B*}=0
\label{eq:sangiovese}
\eq
Note that these equations are the duals of the ones in
(\ref{eq:dog})
and
(\ref{eq:mouse}).
Expanding, we find immediately that
\bq
B= B*   + \left( {d^2 Q\over dB^2}|_B*\right) ^{-{1\over 2}} 4(w-
w*)^{1\over 2} + ....
\label{eq:chianti}
\eq

For the case studied in the previous section with $H_0$
given as in (\ref{deter}) we see from the explicit form given
in
(\ref{eq:pink}) that $B(w)$ indeed has a square root cut
starting at
$w*= \pm {i\over 2\epsilon}$, in agreement with the general
analysis given here.

\section{Non-abelian Central Limit Theorem}

Gauss proved that if we add $K$ random numbers $x_i$,
$i=1,2,...K$,
with $x_i$ taken from the probability distribution $P_i(x_i)$,
then the
normalized sum $s= {1\over \sqrt{K}} \sum_i x_i$ follows the
Gaussian
distribution in the limit $K$ tending to infinity. This result
plays an
important role in physics and mathematics and accounts for
the ubiquitous
appearance of the Gaussian distribution.

What if the variables $x_i$ do not commute? In particular,
suppose that  instead of real numbers $x_i$, we have $N$ by
$N$ random matrices
$\varphi_i$,
$i=1,2,...K$,
taken from from the probability distributions
\bq
P_i(\varphi_i)={1 \over Z_i}{e^{{-N}
trV_i(\varphi_i)}}.
\label{eq:distributioni}
\eq
Does the normalized sum of all these matrices $\varphi_T \equiv
{1\over
\sqrt{K}} \sum_i
\varphi_i$ follow the Gaussian distribution in the large $K$
limit?
Intuitively, it seems that this ought to be the case.
(Note that there exists another commonly
considered class of random matrices, in which the element
of the random
matrices is each taken from a probability distribution (the
same for each
element). For this class, which we refer to as the Wigner
class in our
earlier work \refto{BZ2},
the proposed theorem follows immediately from
the usual abelian central limit theorem. Here we are speaking of the trace
classes defined by (\ref{eq:distributioni}).

As it turns out, it is not difficult to generalize one of the
standard
proofs of Gauss's theorem to matrices. We will give this
proof below.
However, it would seem that the algorithm developed here
for adding random
matrices is almost tailor made to address this question of
whether
$\varphi_T \equiv
{1\over
\sqrt{K}} \sum_i
\varphi_i$ follows the Gaussian distribution. It is mildly amusing to see
how the Wigner semi-circle law emerges naturally.

As we
mentioned earlier,
when we add two random matrices, in general it is
difficult to determine explicitly the resulting $G(z)$ for the
sum
of the two matrices. What we hope for here is that the large
$K$
limit will bring considerable simplification. This is indeed the
case.
To keep the formulas simple,
let us again take $V_i$ to be even. We expect that our
conclusions can be easily
generalized. In this case, $G_i(z)$ is an odd function and
hence $B_i(z)$
is also an odd
function. Thus, we write $B_i(z) = {1\over z} +zb_i(z)$ with
$b_i(z)$ an
even
function. From the law of addition (\ref{eq:addlaw}) given in
this paper,
we learn that the
function $B(z)$ associated with the unknown $G(z)$ is given
by
\bq
B(z)={1\over z}+zb(z)
\label{eq:sumB22}
\eq
where
\bq
b(z)=\sum_{i=1}^K b_i(z)
\label{sum}
\eq
Thus, we obtain the simple
result
that the Green's function $G(z)$ associated with $\sum_i
\varphi_i$ is
determined by
solving
\bq
{1\over G(z)}+b(G(z)) G(z)=z
\label{gauss}
\eq

We are however interested in the normalized sum
$\varphi_T \equiv
{1\over \sqrt
K}\sum_i \varphi_i$. Define the corresponding Green's
function as
$G_T(z)
\equiv
<{1\over N} tr {1\over  z -
 {1\over \sqrt K}\sum_i \varphi_i}>={\sqrt K}G({\sqrt K} z)$.
Rescaling,
we
find that
$G_T(z) $ is determined by
\bq
{1\over G_T(z)}+\left[ {1\over K} \sum_i^K b_i({G_T(z)\over{\sqrt
K}})\right] G_T(z)=z
\label{eq:twidd}
\eq

Note that the argument of $b_i$ in (\ref{eq:twidd}) is
${G_T(z)\over{\sqrt
K}}$.
For finite $K$, (\ref{eq:twidd}) is hopelessly complicated.
However, as $K
\rightarrow
\infty$ we see that it simplifies rather naturally to
\bq
{1\over G_T(z)}+\sigma^2 G_T(z)=z
\label{twidd2}
\eq
with
\bq
\sigma^2 \equiv {1\over K} \sum_i^K b_i(0)
\label{sigma}
\eq
Solving this quadratic equation, we find immediately that
\bq
G_T(z)={1\over 2}\left( z-{\sqrt {z^2 - 4 \sigma^2}}\right)
\label{sol}
\eq
and thus Wigner's semi-circle law for the density
$\rho(\mu)={1\over 2\pi
\sigma^2}{\sqrt {4 \sigma^2 - \mu^2}}$.

What is $b_i(0)$? We note that with $G_i(z) \rightarrow
{1\over z} +
{\sigma_i^2
\over z^3} +.....$ for large $z$ it is easy to show that
$b_i(0)=\sigma_i^2$. Thus, not
only do we obtain the
Wigner
semi-circle law, we learn that $\sigma^2 = {1\over K}
\sum_i^K
\sigma_i^2$. We
should remark that in general $\sigma_i^2$ is not directly
related to the
width of the spectrum of eigenvalues of $\varphi_i$, as the
reader can
easily check by using the explicit formulas given in the
footnote
\refto{foot}.

While we have proved that the density of eigenvalues of
$\varphi_T$
satisfies
the
semi-circle law we cannot yet conclude that the probability
distribution of
$\varphi$
is Gaussian. The reason is that for matrices in the Wigner
class, even when
the
distribution is not Gaussian, the corresponding density of
eigenvalues
still satisfies the
semi-circle law, as is well known. (For a simple proof based
on a
renormalization
group inspired approach, see \refto{BZ2}.) For the trace
class, on the
other
hand, it is
true that if the density satisfies the semi-circle law, then
the
distribution of the
matrices is indeed Gaussian \refto{BIPZ}. However, we see no
reason that
$\varphi_T$ would belong to the trace class, and not to a
more involved
class of probability distributions such as those discussed in
\refto{BZ2}.

As mentioned above, it is not difficult to extend one of the
usual proofs of
the central limit
theorem to the case of matrices. The distribution for the
normalized sum
matrix
$\varphi \equiv {1\over \sqrt K}\sum_i \varphi_i$ is given
by (here we
omit the
subscript $T$)
\bq
P(\varphi)= \left({\prod_i^K} \int d\varphi_i P(\varphi_i)\right)
\delta\left(\varphi - {1\over
\sqrt
K}\sum_i \varphi_i\right)
=\int dt \left( {\prod_i^K} \int d\varphi_i P(\varphi_i)\right) e^{ {i\over
\sqrt
K}\sum_i^K tr
t \varphi_i
} e^{-i tr t\varphi}
\label{eq:sumdist}
\eq
The integral over $\varphi_i$ can be done in the large $K$
limit:
\bq
\int d\varphi_i P(\varphi_i) e^{ {i\over \sqrt K}\sum_i tr t
\varphi_i }
=1 - {1\over 2K}\int d\varphi_i P(\varphi_i) tr t\varphi_i tr
t\varphi_i +
O\left( {1\over K^2}\right)
=1 - {\sigma_i^2 \over 2KN} tr t^2  +
O\left( {1\over K^2}\right)\\
\label{eq:quad}
\eq
where we have defined $\sigma_i^2 = \int d\varphi_i
P(\varphi_i) {1\over
N} tr
\varphi_i^2$.  Reexponentiating and integrating over $t$ we
obtain the
desired
result that
$P(\varphi)$ is proportional to $e^{- {N\over 4 \sigma^2}   tr
\varphi^2 }
$.

\section*{Acknowledgement}

I thank the organizers of the workshop on ``Operator
Algebra Free
Products
and Random Matrices" held at the Fields Insitute, March
1995, for inviting
me to talk about my work. I had the opportunity there to
learn  from
mathematicians, in particular U. Haagerup and D.
Voiculescu, about recent
developments. I am especially grateful to M.  Douglas for
many instructive
discussions and for telling me how to translate from the
language used by
mathematicians to the language used by physicists. Finally,
I would like to
thank E. Br\'ezin, my collaborator over the last few years,
for countless
helpful discussions about this and other topics in random
matrix theory. The results of sections 4 - 6
were worked out
together with him.
This work was supported in part by the National Science
Foundation under Grant No. PHY89-04035.

\section*{Figure Captions}

{\bf Fig 1.} Feynman rules: (a) quark propagator, (b) gluon
propagator, (c)
quark
gluon vertex, (d) gluon interaction, illustrated here with a
$g \varphi^4$
vertex.

\noindent {\bf Fig 2.} Quark propagator and one-particle-
irreducible self
energy.

\noindent {\bf Fig 3.} Quark self energy: the gluons shown
explicitly are
all
of type 1.
There are of course also type 2 gluons inside the quark
propagator $G$.

\noindent {\bf Fig 4.} A class of diagrams not included in
(3d).


\references

\refis{WIG} E. Wigner, Can.~Math.~Congr.~Proc.~
(University of Toronto Press) p. 174, and other papers
reprinted in Porter, op. cit.

\refis{POR} C.E. Porter, Statistical Theories of
Spectra: Fluctuations (Academic Press, New York,
1965).

\refis{MEH} M.L. Mehta, Random Matrices
(Academic
Press, New York, 1991).

\refis{JER} See for instance, Two Dimensional
Quantum Gravity
and Random Surfaces, edited by D.~J.~
Gross and T.~Piran (World Scientific, Singapore, 1992).

\refis{BIPZ} E. Br\'ezin, C. Itzykson, G. Parisi, and J.B.
Zuber, Comm.~Math.~Phys. 59 (1978) 35.

\refis{BZ1} E. Br\'ezin and A. Zee, Nucl.~Phys.~B 402(FS)
(1993) 613.

\refis{amb} J. Ambj\o rn, ``Quantization of geometry," in Les
Houches 1990, edited by J.~Dalibard et al, J.~Ambj\o rn, J.~Jurkiewicz,
Yu.~M.~Makeenko, Phys.~Let.~B 251 (1990) 517.

\refis{bee} C.W.J. Beenakker, Nucl.~Phys.~B 422 [FS] (1994) 515.

\refis{danna} J. D'Anna, E. Br\'ezin, and A. Zee, Nucl.~Phys. [FS] 
in print, 1994.

\refis{for} P. J. Forrester, Nucl.~ Phys.~B 435 [FS] (1995) 421.

\refis{koba} T. S. Kobayakawa, Y. Hatsugai, M. Kohmoto,
and A. Zee, Phys.~Rev.~E 51 (1995) 5365.

\refis{BZ2} E. Br\'ezin and A. Zee, Compt.~Rend.~Acad.~Sci.
(Paris) t.317 II (1993) 735.

\refis{BZ3} E. Br\'ezin and A. Zee, Phys.~Rev.~E 49 (1994) 2588.

\refis{BHZ} E. Br\'ezin, S. Hikami and A. Zee, Phys.~Rev.~E 51 (1995) 5442.

\refis{voi} D. V. Voiculescu, K. J. Dykema, and A. Nica, Free Random 
Variables (AMS, Providence, RI, 1992).

\refis{hag} U. Haagerup, private communication and to
be published.

\refis{waterloo} Lectures at the workshop on ``Operator
Algebra Free
Products and Random Matrices," Fields Insitute (March
1995) To appear in
the Proceedings.

\refis{douglas} M. Douglas, Rutgers preprint, hep-
th/9409098 (1994).

\refis{gross} R. Gopakumar and D. J. Gross,  Princeton
preprint PUPT-1520 (1994).

\refis{lattice} E. Br\'ezin and A. Zee, Nucl.~Phys.~B 441 [FS] (1995) 409.

\refis{bz4} E. Br\'ezin and A. Zee, Nucl.~Phys.~B 424 [FS] (1994) 435.

\refis{li} M. Douglas and M. Li, Rutgers preprint (1995).

\refis{lattice} E. Br\'ezin and A. Zee, Nucl.~Phys.~B 441 [FS] (1995) 409.

\refis{girvin} C. B. Hanna, D. P. Arovas, K. Mullen, and S.
M. Girvin, Indiana preprint (1994) [cond-mat 9412102].


\refis{hikami} S. Hikami, M. Shirai, and F. Wegner, Nucl.~ Phys.~B  408 
(1993) 415.

\refis{hz} S. Hikami and A. Zee, Tokyo-Santa Barbara
preprint (1995) [cond-mat/9504014] Nucl.~Phys. [FS] in print.

\refis{eyn} B. Eynard, Nucl.~Phys. [FS] in print, (1994).

\refis{thoo} G. 't Hooft, Nucl.~Phys. B72 (1974) 461.

\refis{PAS} L.A. Pastur, Theo.~Math.~Phys. 10 (1972) 67.

\refis{foot} For the sake of completeness, even though we
won't need these explicit formulas in the text, let us record
(and for the sake of simplicity, with $V$ an even function) that
for $V(\varphi)=\sum_{k=1}^p {1\over 2k}g_k \varphi^{2k}$
we have
\bq
P(z)={1\over 2}\sum_{k=1}^p g_k \sum_{n=0}^{k-1}
{(2n)!\over (n!)^2}
({a^2\over 4})^n \lambda^{2k-2n-2}
\eq
and
\bq
{1\over 2}\sum_{k=1}^p g_k {(2k)!\over (k!)^2}({a^2\over
4})^k =1
\eq
In particular, for $V={1\over 2}\varphi^2+{g\over
4}\varphi^4$ we have
\bq
G(z)={1\over 2}[z+gz^3-(1+{a^2 g\over 2}+g z^2) {\sqrt
{z^2-a^2}}]
\eq
and
$a^2={2\over 3g}(\sqrt {1+12g}-1)=4(1-3g+18g^2+....)$.

\refis{passing} It is perhaps worth noting that $B_2$
satisfies an equation
similar to (\ref{eq:sigma}), namely
\bq
P^2(B_2)(B_2^2 - a^2)=(V'(B_2)-2z)^2
\eq

\refis{addgauss} Using dimensional analysis we see
immediately that for
$V(\varphi)={1\over 2}m^2 \varphi^2$ the Blue's function
$B(z)={z \over
m^2}+{1\over z}$. Then the addition law in (\ref{eq:addlaw})
implies the usual
Gaussian law of
addition $m^{-2}_{1+2}=m^{-2}_1 + m^{-2}_2$.

\endreferences
\bye